# Characterization of 3D surface topography in 5 axis milling

Yann Quinsat[1*], Sylvain Lavernhe[1], Claire Lartigue[1,2]

[1] LURPA - ENS de Cachan – Université Paris Sud 11
61 avenue du Président Wilson
94235 Cachan cedex – France
fax: +33 1 47 40 22 20
name@lurpa.ens-cachan.fr

[2] IUT de Cachan – Université Paris Sud 11
9 avenue de la Division Leclerc
94234 Cachan cedex – France

[*] corresponding author
tel: +33 1 47 40 22 13

**Abstract**

Within the context of 5-axis free-form machining, CAM software offers various ways of tool-path generation, depending on the geometry of the surface to be machined. Therefore, as the manufactured surface quality results from the choice of the machining strategy and machining parameters, the prediction of surface roughness in function of the machining conditions is an important issue in 5-axis machining. The objective of this paper is to propose a simulation model of material removal in 5-axis based on the N-buffer method and integrating the Inverse Kinematics Transformation. The tooth track is linked with the velocity giving the surface topography resulting from actual machining conditions. The model is assessed thanks to a series of sweeping over planes according to various tool axis orientations and cutting conditions. 3D surface topography analyses are performed through the new areal surface roughness parameters proposed by recent standards.

## 1. Introduction

In the field of free-form machining, CAM software offers various machining strategies depending on the geometry of the surface to be machined. The surface quality results from the choice of the machining strategy and corresponding parameters (tool inclination, feed per tooth, cutting speed, radial depth of cut). Resulting machining time, productivity and geometrical surface quality directly depend on these parameters. In 5 axis machining, axis kinematical capacities as well as specific NC treatments alter tool trajectory execution, leading to changes in actual local feedrates. Moreover, as the tool axis orientation generally varies during machining, the resulting surface pattern can be affected [1]. The prediction of the 3D surface topography according to the machining conditions is an important issue in 5-axis machining to correctly achieve process planning and to link resulting surface patterns with part functionality.

### 1.1 Surface topography description

With the advances in 3D measuring systems, it is now possible to measure machined surface patterns with enough accuracy [2,3,4] although there is no standard traceability [5]. A draft standardized project [ISO 25178-2] developed by the ISO Technical Committee 213 working group 16, proposes the definition of areal parameters as an extension of the well-known profile parameters [6] [7]. However, only a few studies try to link the surface roughness with surface requirements via areal surface roughness parameters. For friction in servo hydraulic assemblies, negative Skewness and the lowest Kurtosis values as well as the highest valley fluid retention index are found to have the lowest frictional characteristics [8]. The functionality of automotive cylinder bores is partially characterized by oil consumption and blow-by. In this specific case, it is more significant to consider $Sq$, $Sk$, $Svk$, $Sds$, $Sbi$ to describe oil consumption and $Sv$, $Svi$ for blow-by [9]. Concerning the fatigue limit, authors prefer to refer to $Sq$, $Std$ and $Sal$ [10]. Due to the lack of information concerning the influence of roughness parameters on surface requirement, a description of the 3D pattern obtained after surface machining is essential to bring out the influence of machining parameters on surface topography, and to afterwards link surface roughness with functional requirements.

### 1.2 Surface topography prediction

In the literature, few formalized studies exist which aim at linking the surface topography with the machining strategy parameters [11]. Two standpoints can be adopted: the experimental standpoint and the theoretical standpoint. Based on surface topography measurements, most of the experimental methods attempt to establish the link between the feedrates, the machining direction, the tool orientation and the 3D topographies. Unfortunately, results are only qualitative; only a few of them clearly express the relationship between the machining strategy parameters and the surface topography [12], [13]. Adopting the theoretical standpoint, Kim described the texture obtained in ball-end milling from numerical simulations only accounting for the feedrate influence [14]. Bouzakis focused on the motion of the cutting edge. The author highlights the influence of the tool orientation, the transversal step and the feedrate on the machined surface quality [15]. Toh supplements this work by defining the best direction to machine an inclined plane [16]. In a previous work, we proposed to link the machining strategy in 3-axis ball-end milling with a 3D surface roughness parameter and to optimize the machining direction according to this parameter [17]. Kim proposed to simulate the 3D topography obtained in 5 axis milling using a filleted-ball end tool. The envelope of the tool movement is modelled by successive tool positioning according to the feed per tooth.

Due to difficulties in measuring the surface topography for complex shapes, the need for models or simulations for predicting the machined 3D surface topography is real. However, if most literature works enhance the major role of the federate, the context of high speed machining is seldom considered. Actually, in multi-axis high speed machining the computation of the inverse kinematic transformation and the synchronisation of the rotational axes with the translation ones impact the respect of the programmed feedrate which does not remain constant during machining. Therefore, it seems essential to integrate those local federate variations in a prediction model of 3D surface topography obtained in multi-axis high-speed machining.

In this paper, a theoretical approach is proposed to predict the 3D surface topography obtained in 5-axis milling with a filleted-ball end cutter tool integrating actual feedrate evolution.

Actual feedrate evolution is obtained thanks to a kinematical predictive model which accounts for the local variations of the velocity due to multi-axis high speed machining [18]. The modelling of the cutting process is only geometrical; material pull out is not consider here. The proposed model applies for complex surfaces for which the topography measurement is generally difficult. The topography prediction relies on the well-known N-buffer simulation method [19].

Based on simulations, the study finally aims at formalizing the influence of the machining parameters (feed per tooth, tool inclination, maximal scallop height allowed) on the 3D surface topography. For this purpose, the topography is characterized using the areal surface parameters. An attempt is made to propose links between areal surface parameters and the parameters of the machining strategy.

## 2. 3D Surface topography in 5 axis machining

Material removal simulation relies on the well-known N-buffer method [19]. The main difficulty is the integration of the effects linked to 5-axis machining within a context of high velocities. Indeed, the use of the two additional rotational axes leads to two main difficulties during trajectory execution: the computation of the Inverse Kinematical Transformation in real time to define set points corresponding to tool postures, and the synchronization of the rotational axes with the translational ones [18]. Moreover, due to kinematical axis limits, axis velocities may vary leading to feedrate fluctuations which can alter the 3D pattern. In the proposed approach, the prediction of the surface topography takes advantage of a model of velocity prediction developed in a previous work which gives a good estimation of the local feedrate of the tool-teeth [18].

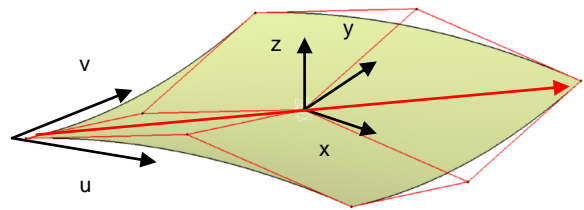

Fig. 1: Hyperbolic paraboloïd

To illustrate this purpose, do consider the example of the surface presented in figure 1. The surface, a hyperbolic paraboloïd with a double curvature, is machined along its rules with a filleted-end tool ($R$=5mm, $r$=1.5mm), considering a tool inclination of 1°(tilt angle = 1°, see figure 4). During machining, the surface curvature involves a combined movement of all the 5 axes. The programmed feedrate is set to 5m/min.Using the predictive velocity model, the calculation of the feedrate all trajectory long is carried out [1]. Figure 2 presents the evolution of the local feedrate for the machining of the trajectory at the middle of the surface (red arrow in figure 1). Simulated values as well as measured ones are reported.

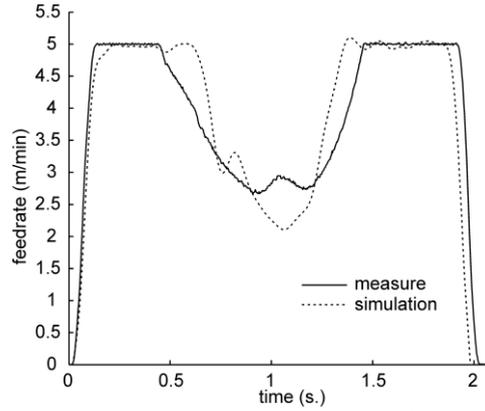
Fig. 2: Simulated and measured feedrates

As it can be observed, whether for the simulation as for the measurement, the programmed feedrate is only reached at the beginning and at the end of the trajectory; velocity is strongly decreased at the middle of the trajectory. However, some differences between simulated and measured values are noticeable: although the velocity decreasing is correctly predicted by simulation, deceleration is faster and occurs later. Nevertheless, simulation gives a good estimate of the feedrate, and thus of the local feed per tooth. Therefore, as actual cutting conditions can be known, a more precise simulation of the 3D surface topography is now possible.

The simulation requires the modelling of the surface, the modelling of the tool geometry and the definition of the actual tool trajectory [1]. The surface is sampled by a grid of points defined in a (XY) plane. A line, parallel the local surface normal, is associated to each point of the grid, thus defining a line-net. This line-net is truncated by the cutter tool according to the actual tool trajectory, and the remaining part of the line-net defines the 3D topography of the machined surface.

For its part, the tool is supposed to be rigid and measured by optical means. The complete tool geometry is approximated by a local meshing, i.e. the cutting edge as well as the tool flank face. Only active cutting edges are considered. To ensure a correct approximation of the tool surface, the meshing is performed with a chord error equal to 0.1μm.

Concerning the tool trajectory, the proposed method integrates actual local feedrates calculated using the prediction model (figure 3). More generally, the tool trajectory is defined in the part coordinate system (PCS) by a set of tool postures. Considering the velocity prediction model, local feedrates $V_f^i$ can be calculated for each tool posture. Therefore, a tool posture belonging to the trajectory is defined by $\{X_p^i, Y_p^i, Z_p^i, I^i, J^i, K^i, V_f^i\}$, where $(X_p^i, Y_p^i, Z_p^i)$ are the coordinates of the tool tip and $(I^i, J^i, K^i)$ are the axis tool direction components.

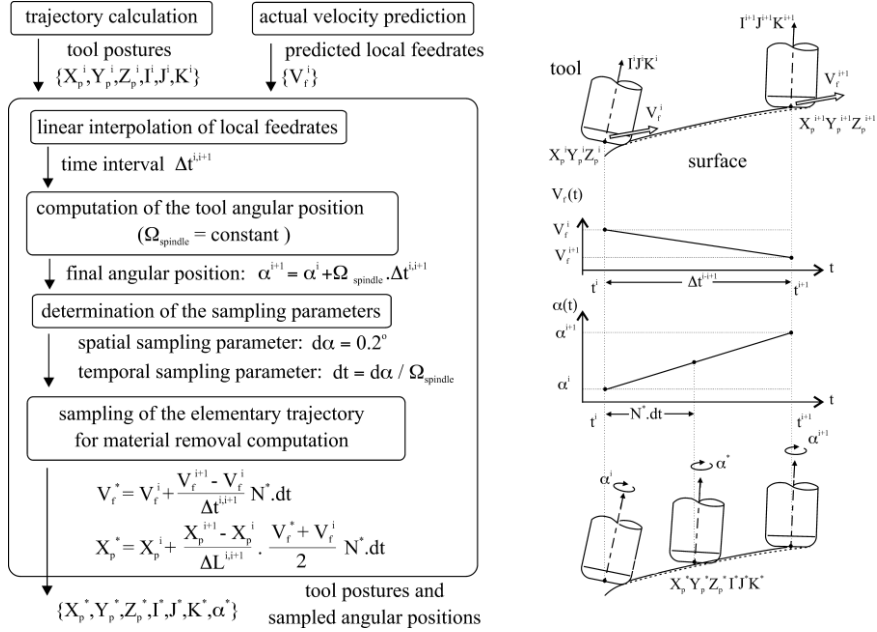
Fig. 3: Local trajectory calculation in 5-axes

The displacement of the tooltip between two postures is defined as:
$$\Delta L^{i,i+1} = \sqrt{\left(X_p^{i+1} - X_p^i\right)^2 + \left(Y_p^{i+1} - Y_p^i\right)^2 + \left(Z_p^{i+1} - Z_p^i\right)^2} \qquad \text{Eq. (1)}$$

Hence, based on the linear interpolation of the feedrate the time interval separating two tool postures is calculated as follows:
$$\Delta t^{i,i+1} = \frac{\Delta L^{i,i+1}}{\dfrac{V_f^{i+1} - V_f^i}{2}} \qquad \text{Eq. (2)}$$

Supposing the rotational velocity of the spindle $\Omega_{spindle}$ to be equal to the programmed one, $\{\dot{\alpha}\}$, the angular positions of the tool axis are given by:
$$\alpha^{i+1} = \alpha^i + \Omega_{spindle} \cdot \Delta t^{i,i+1} \qquad \text{Eq. (3)}$$

The elementary trajectory defined between two tool postures is afterwards sampled considering a fixed step, $d\alpha$:
$$\alpha^* = \alpha^i + N^* \cdot d\alpha \qquad \text{Eq. (4)}$$

where $N^*$ is an integer belonging to the interval $[1, \text{floor}(\dfrac{\alpha^{i+1} - \alpha^i}{d\alpha})]$.

Therefore, the temporal sampling parameter is calculated using the following equation:
$$dt = \frac{d\alpha}{\Omega_{spindle}} \qquad \text{Eq. (5)}$$

For each sampling point ($N^*$), the local feedrate is thus expressed by:
$$V_f^* = V_f^i + \frac{V_f^{i+1} - V_f^i}{\Delta t^{i,i+1}}(N^* \cdot dt) \qquad \text{Eq. (6)}$$

This yields to the calculation of the sampled tool locations along the elementary trajectory:
$$X_p^* = X_p^i + \left(\frac{X_p^{i+1} - X_p^i}{\Delta L^{i,i+1}}\right) \cdot \frac{V_f^* + V_f^i}{2} N^* \cdot dt \qquad \text{Eq. (7)}$$

Finally, the simulated machined surface is obtained by computing the intersections between the normal lines defining the part and the tool for each configuration $\{X_p^*, Y_p^*, Z_p^*, I^*, J^*, K^*, \alpha^*\}$.

## 3. Model assessment

### 3.1 Model assessment for plane surfaces

The model is assessed by comparing 3D surface topographies obtained by simulations to actual measured ones for different types of part. The first validation concerns the milling of a series of planes considering variable machining strategy parameters: the tool axis orientation, the programmed feedrate ($V_f$) and the maximal scallop height allowed ($h_c$) (Table 1). In the proposed experiments, the tool orientation is defined by the tilt angle ($\theta_t$) and the yaw angle ($\theta_n$). A complete experimental design is performed, considering 2 levels per factor, except for the yaw angle, for which 3 levels are considered. The machining is performed on a 5-axis machine tool using a filleted-end milling tool ($R$=5mm, $r_c$=1.5mm) with a unique tooth in order to control the tooth geometry which contributes to the final imprint.

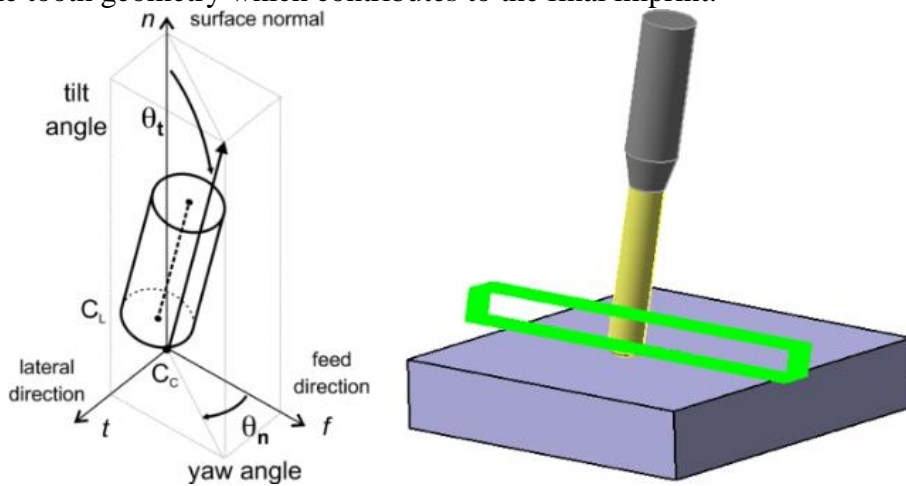

Fig.4: Definition of the experimental design

| Yaw angle ($\theta_n$ °) | | | Tilt angle ($\theta_t$ °) | | Scallop height ($h_c$ mm) | | Feedrate $V_f$ (m/min) | |
|---|---|---|---|---|---|---|---|---|
| 0 | 20 | 40 | 1 | 10 | 0,005 | 0,001 | 2 | 4 |

Tab. 1: Experimental parameters

After machining, resulting surface topographies are measured using a coherence scanning interferometer. To characterize the obtained pattern, 3D parameters define in the draft standard [ISO 25178-2] are used. Although a complete experimental design has been performed only a few cases are reported in table 2. Nevertheless, for all the cases, simulated patterns as well as defect magnitudes match the measured ones. Some small deviations can be observed, probably due to the cutting process or/and the effect of the actual tool geometry.

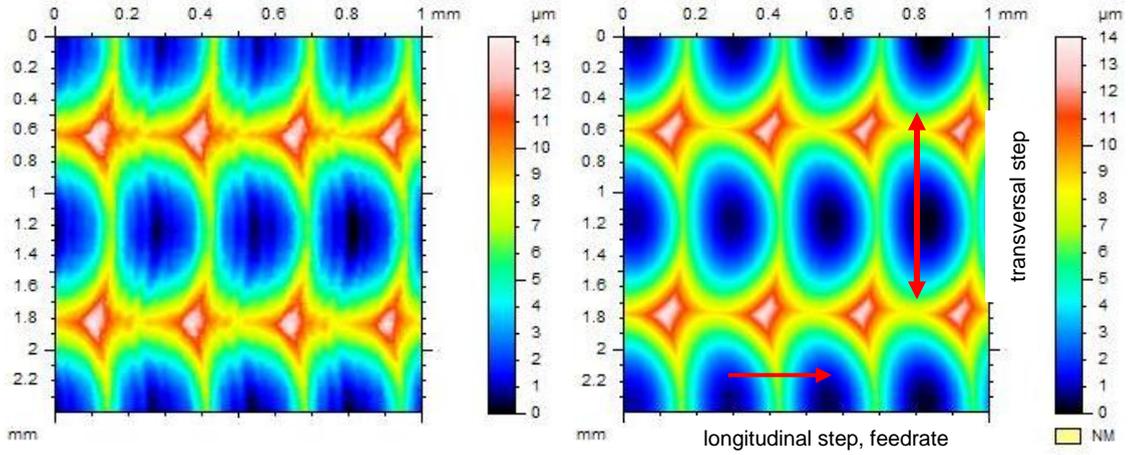

Fig. 5: Measured and simulated pattern for case 1

| case n° | $\theta_n$ (°) | $\theta_t$ (°) | $h_c$ (mm) | $v_f$ (m/min) | Sz (μm) | | Sa (μm) | | Trans. step (mm) | | Long. step, $f_z$ (mm) | |
|---|---|---|---|---|---|---|---|---|---|---|---|---|
| 1 | 0  | 1  | 0,005 | 2 | 6.99  | 5.58 | 1.27 | 1.21 | 2.76 | 2.63 | 0.13 | 0.14 |
| 2 | 0  | 1  | 0,005 | 4 | 9.17  | 9.24 | 1.46 | 1.66 | 2.71 | 2.62 | 0.26 | 0.27 |
| 3 | 0  | 10 | 0,01  | 4 | 14.10 | 15.7 | 2.47 | 2.56 | 1.21 | 1.18 | 0.26 | 0.27 |
| 4 | 20 | 10 | 0,01  | 4 | 6.54  | 5.63 | 0.96 | 1.05 | 0.46 | 0.43 | 0.26 | 0.27 |

Tab. 2: Comparison between measured and simulated patterns

The experimentation enhances the influence of the feedrate on the 3D parameters characterizing the surface topography. As in a previous study, an attempt is made to link $Sz$, the maximum height of the surface with the machining parameters [17]. For this purpose, an analytical model is defined from the expression of the effective cutting radius. Let us consider an approximation of the effective cutting tool radius by the equivalent radius ($R_{eq}$) defined at the contact point in function of the tool axis orientation (defined by the couple ($\theta_n$, $\theta_t$)) [20]:

$$R_{eq} = \frac{r(R + r.\sin(\theta_t))}{R.\sin(\theta_t).(\cos(\theta_n))^2 + (R + r.\sin(\theta_t)).(\cos(\theta_n))^2} \qquad \text{Eq. (8)}$$

The analytical model highlights the influence of the scallop height ($h_c$) compared to the feedrate ($f_z$) for different tool orientations. Therefore, if $r$ is the corner radius of the tool, $S_z$ is estimated by:

$$Sz = f_z^2/8.r \qquad \text{if } r > \sqrt{8.h_c.R_{eq}} \qquad \text{Eq. (9)}$$
$$Sz = h_c + f_z^2/8.r \qquad \text{otherwise}$$

In the case of the experimental design previously developed, this approximation gives a mean error of 1.58 μm for the simulation and 1.67 μm for the measurement. The standard deviation is respectively 1.22 μm for the simulated values, and 1.26 μm for the measured ones. To summarize, the analytical model gives a relationship useful to link the tool axis orientation, and the machining parameters to the maximum height deviation *Sz*, in the case of a plane surface.

### 3.2 Model assessment for complex surfaces

One main interest of the model is that it applies for the machining of complex surfaces. Indeed, due to measuring system capacities, measurement on complex surfaces is generally difficult as regards its curvature. The model is applied to the paraboloïd. Considering the red tool path defined in figure 6, the machining is performed according to the following machining conditions: tool orientation defined by ($\theta_t=1°$, $\theta_n=0°$); distance between passes equal to 5mm; programmed feedrate equal to 5 m/min.

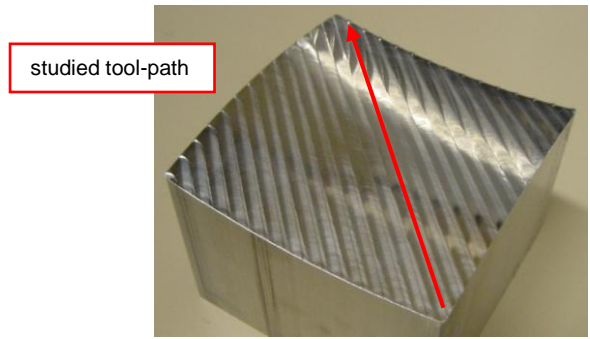

Fig. 6: Machined part

The 3D topography obtained after machining is simulated considering the predicted velocities (figure 2). As it was highlighted, the velocity is strongly decreased at the middle of the trajectory due to the kinematics limits and only reaches half the programmed feedrate. The simulated pattern is reported in figure 7 (right). On the other hand, the part is measured using the chromatic sensor. The measured topography reported in figure 7 is close to the simulated one. Defect magnitudes as well as the patterns are similar. Differences are probably due to the actual cutting phenomenon. Nevertheless, such differences are small enough to assess the model relevancy for predicting 3D surface topography of complex surfaces machined in 5-axis milling.

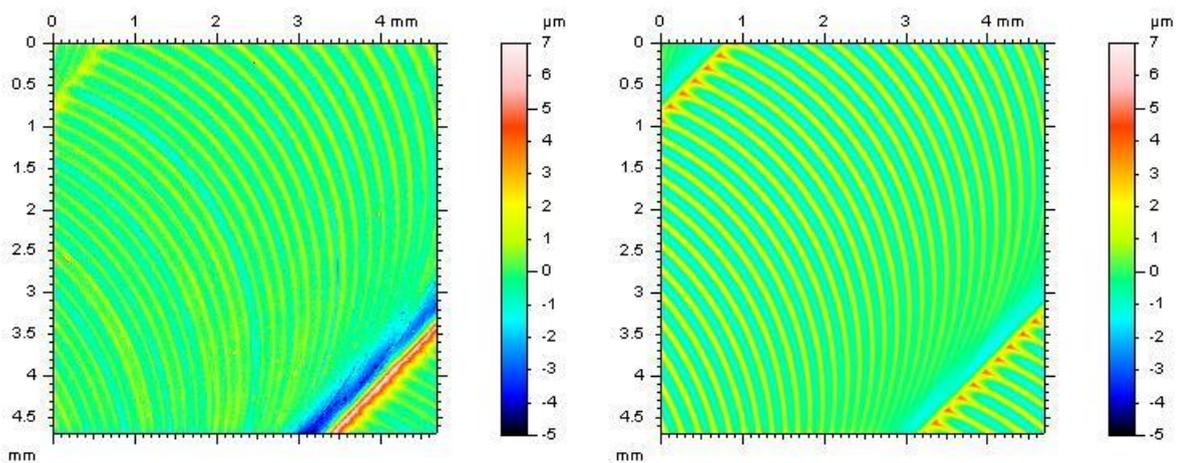

Fig. 7: Measured (left) and predicted (right) topographies for the paraboloïd

Hence, the analysis of influent parameters on the surface topography can be conducted through simulations only.

## 4. 3D surface topography parameters

The complete experimental design is also conducted through simulations, considering experimental parameters defined in table 2. As previously discussed, the feedrate is an essential parameter, as it actually conditions the 3D pattern (Figure 8). Modifications of local feedrate during machining may affect the 3D surface finish.

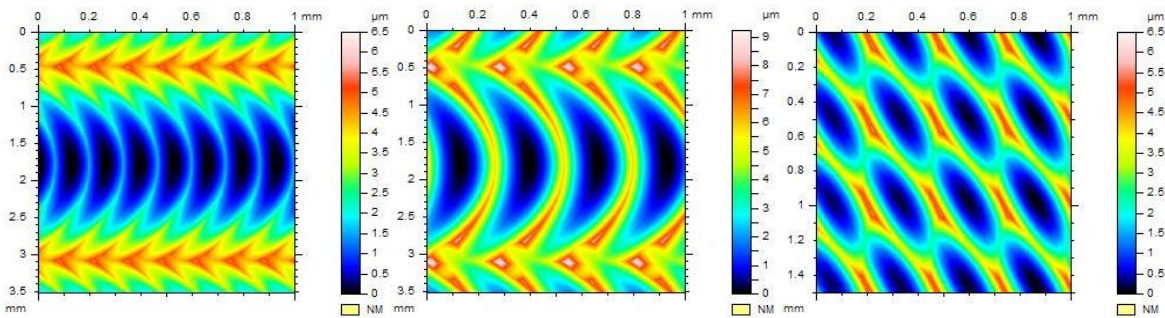

Fig. 8: Simulated patterns (from left to right cases 2, 3 and 4 – Table 2)

Usually, the maximum scallop height allowed is one of the most used parameters in CAM software to define the 3D surface topography. As shown in figure 8 (case 4), a non null yaw angle provides a pattern for which the notion of cusp has no more significance. This enhances the major influence of the tool inclination in surface patterns resulting from 5-axis machining. According to previous works aiming at linking areal surface roughness parameters with part functionality [8,9,10], the analysis of the experimental is only conducted for the parameters the most significant for fatigue and friction applications. These parameters are classified in function of their family: amplitude parameters $Sa$, $Sq$, $Sv$, $Ssk$, $Sku$ and spatial parameters $Sal$, $Sds$, $Std$. Results relative to our experimentation are given in table 3. The effect of each factor is calculated as follows:

$$S_i = \text{Mean}_i + \sum_j \text{Effect}_j \cdot \begin{bmatrix} 1 \\ -1 \end{bmatrix} \qquad \text{Eq. (10)}$$

Where j = yaw, tilt, scallop height, feedrate

Concerning $Std$, the screw angle seems the more relevant influent parameter. This is consistent with pattern observations. In fact the marks left by the tooth are oriented according to this angle. Due to the modification of the effective cutting radius, the screw angle has also a significant effect on $Sal$, $Std$, $St$ and $Sv$. On the opposite, its effect is quite negligible on kurtosis or skewness values. The tilt angle is the most significant parameter for $Sq$, with a little effect on $Sal$. However, it does not influence the texture direction $Std$. Results emphasise that feedrate is more influent than the maximum scallop height on the studied 3D parameters. Particularly for the distribution of peaks ($Sds$), the feedrate is the most significant parameter.

| Parameter | | | Mean | Effect | | | |
|---|---|---|---|---|---|---|---|
| | | | | Yaw | Tilt | Scallop height | Feedrate |
| Amplitude | Maximum height of the surface | Sz (μm) | 4,79 | -2,34 | 0,56 | -0,82 | 1,27 |
| | Arithmetic deviation of the surface | Sa (μm) | 0,92 | -0,45 | 0,19 | 0,1 | 0,16 |
| | Root-Mean-Square Deviation of the Surface | Sq (μm/μm) | 1,10 | -0,32 | -0,57 | 0,12 | 0,24 |
| | Kurtosis of Topography Height Distribution | Sku (no unit) | 2,33 | 0,01 | -0,04 | -0,17 | 0,06 |
| | Skewness of Topography Height Distribution | Ssk (no unit) | 0,47 | 0,00 | -0,07 | -0,05 | -0,02 |
| Volume | Valley Void Volume of the Surface | Sv (μm) | 1,71 | -0,81 | 0,25 | -0,28 | 0,47 |
| Spatial | Density of Summits of the Surface | Sds (pics/mm$^2$) | 606,38 | 132,87 | -115,13 | -31,68 | -153,50 |
| | The Fastest Decay Autocorrelation Length | Sal (mm) | 0,15 | -0,12 | -0,07 | -0,03 | -0,04 |
| | Texture Direction of the Surface | Std (°) | -21,80 | -13,33 | 0,03 | -0,44 | -0,70 |

Tab. 3: Mean values of the effects

Nonetheless, previous observations must be modulated as influences of the machining strategy parameters are close to each other. In addition, interactions between parameters have not been studied in this work. Indeed, the scallop height, the yaw and the tilt angles are linked by the transversal step calculation. This actually binds respective influences.

## 5. Conclusion

The objective of the present paper is to propose a method for characterizing 3D topographies of complex machined surfaces. For this purpose, a simulation model of material removal in 5-axis milling is developed and assessed. As in 5-axis machining, velocities are non uniform during machining and vary linked to kinematical limits, the model is coupled to a velocity prediction model allowing the determination of actual local feeds per tooth. Simulations, compared with measurements, clearly enhances that variable local federates along a trajectory affect the resulting pattern. On the other hand, the effect of machining strategy parameters such as tool inclination and maximum scallop height allowed are investigated thanks to the topography simulation model. The pattern characterization is performed via areal parameters with an attempt to link them with machining strategy parameters. Simulations bring out that depending on the areal parameter chosen, one of the machining parameter is determinant for the surface quality. In particular, the influence both angles defining the tool inclination is significant. Nevertheless, the final objective of the present work is to propose a method for the choice of machining strategy parameters according to the machined surface function. In this direction, an important work remains to link areal surface parameters to part functionality.